\documentclass[floatfix,prd,epsfig,nofootinbib,superscriptaddress,onecolumn,amssymb]{revtex4}

\usepackage{slashed}
\usepackage{slashed}
\usepackage{graphicx,color}
\usepackage{epsfig}
\usepackage{subfigure}
\usepackage{epsfig}
\usepackage{dcolumn}
\usepackage{bm}
\usepackage{color}

\def\lsim{\mathrel{\rlap{\lower4pt\hbox{\hskip1pt$\sim$}}
    \raise1pt\hbox{$<$}}}         
\def\gsim{\mathrel{\rlap{\lower4pt\hbox{\hskip1pt$\sim$}}
    \raise1pt\hbox{$>$}}}         
    
    \newcommand{\nc}{\newcommand}  

\nc{\beq}{\begin{equation}}  
\nc{\eeq}{\end{equation}}  
\nc{\beqa}{\begin{eqnarray}}  
\nc{\eeqa}{\end{eqnarray}}  
\nc{\bea}{\begin{eqnarray}}  
\nc{\eea}{\end{eqnarray}}  
\nc{\ra}{\rightarrow}  
\nc{\slsh}{\slash\hspace*{-0.22cm}}

\def\Re{{\cal R \mskip-4mu \lower.1ex \hbox{\it e}\,}}
\def\Im{{\cal I \mskip-5mu \lower.1ex \hbox{\it m}\,}}
\def\be{\begin{equation}}
\def\ee{\end{equation}}
\def\bea{\begin{eqnarray}}
\def\eea{\end{eqnarray}}
\def\bit{\begin{itemize}}
\def\eit{\end{itemize}}
\nc{\eref}[1]{(\ref{#1})}
\nc{\Eref}[1]{Eq.~(\ref{#1})}

\nc{\vev}[1]{ \left\langle {#1} \right\rangle }
\nc{\bra}[1]{ \langle {#1} | }
\nc{\ket}[1]{ | {#1} \rangle }
\nc{\fb}{\,{\rm fb}^{-1}}
\nc{\ev}{{\rm eV}}
\nc{\kev}{{\rm keV}}
\nc{\Mev}{{\rm MeV}}
\nc{\gev}{{\rm GeV}}
\nc{\tev}{{\rm TeV}}
\nc{\mev}{{\rm MeV}}

\def\ee{e^+e^-}

\def\msb{{\bar{\ssstyle M \kern -1pt S}}}

\begin{document}

\title{Determination of the  Free Neutron Lifetime}

\author{J.~David~Bowman}
\affiliation{Oak Ridge National Laboratory, Oak Ridge, TN}

\author{L.~J.~Broussard}
\affiliation{Los Alamos National Laboratory, Los Alamos, NM}

\author{S.~M.~Clayton}
\affiliation{Los Alamos National Laboratory, Los Alamos, NM}

\author{M.~S.~Dewey}
\affiliation{National Institute of Standards and Technology, Gaithersburg, MD}

\author{N.~Fomin}
\affiliation{University of Tennessee, Knoxville, TN}

\author{K.~B.~Grammer}
\affiliation{University of Tennessee, Knoxville, TN}

\author{G.~L.~Greene\footnote{Corresponding Author}}\email{ggreene@utk.edu}
\affiliation{University of Tennessee, Knoxville, TN}
\affiliation{Oak Ridge National Laboratory, Oak Ridge, TN}

\author{P.~R.~Huffman}
\affiliation{North Carolina State University, Raleigh, NC}

\author{A.~T.~Holley}
\affiliation{Tennessee Technological University, Cookeville, TN}

\author{G.~L.~Jones}
\affiliation{Hamilton College, Clinton, NY}

\author{C.-Y.~Liu}
\affiliation{Indiana University/CEEM, Bloomington, IN}

\author{M.~Makela}
\affiliation{Los Alamos National Laboratory, Los Alamos, NM}

\author{M.~P.~Mendenhall}
\affiliation{National Institute of Standards and Technology, Gaithersburg, MD}

\author{C.~L.~Morris}
\affiliation{Los Alamos National Laboratory, Los Alamos, NM}

\author{J.~Mulholland}
\affiliation{University of Tennessee, Knoxville, TN}

\author{K.~M.~Nollett}
\affiliation{University of South Carolina, Columbia, SC}
\affiliation{San Diego State University, San Diego, CA}

\author{R.~W.~Pattie, Jr.}
\affiliation{Los Alamos National Laboratory, Los Alamos, NM}

\author{S.~Penttil\"{a}}
\affiliation{Oak Ridge National Laboratory, Oak Ridge, TN}

\author{M.~Ramsey-Musolf}
\affiliation{University of Massachusetts, Amherst, MA}

\author{D.~J.~Salvat}
\affiliation{Indiana University/CEEM, Bloomington, IN}
\affiliation{Los Alamos National Laboratory, Los Alamos, NM}

\author{A.~Saunders}
\affiliation{Los Alamos National Laboratory, Los Alamos, NM}

\author{S.~J.~Seestrom}
\affiliation{Los Alamos National Laboratory, Los Alamos, NM}

\author{W.~M.~Snow}
\affiliation{Indiana University/CEEM, Bloomington, IN}

\author{A.~Steyerl}
\affiliation{University of Rhode Island, Kingston, RI}

\author{F.~E.~Wietfeldt}
\affiliation{Tulane University, New Orleans, LA}

\author{A.~R.~Young}
\affiliation{North Carolina State University, Raleigh, NC}

\author{A.~T.~Yue}
\affiliation{National Institute of Standards and Technology, Gaithersburg, MD}

\maketitle

\section{Executive Summary}
Neutron beta decay is an archetype for all semileptonic charged-current weak nuclear processes. As a result, an accurate determination of the parameters that describe neutron decay is critical for the detailed understanding of a wide variety of nuclear processes. In cosmology, the neutron lifetime determines weak interaction rates and therefore the helium yield of big bang nucleosynthesis (BBN). The neutron lifetime, along with neutron decay correlations, nuclear and kaon decay data, can be used to test the unitarity of the CKM matrix and probe physics beyond the standard model. Such tests are highly complementary to information that is anticipated from the LHC.

Given its importance, it is disturbing that the most accurate determinations of the neutron lifetime are discrepant. The neutron lifetime has been measured by a decay-in-flight method known as the ``beam method'' and a neutron confinement method known as the ``bottle method''~\cite{few11}.  While there is consistency among the beam measurements and among the bottle measurements, the two sets of experiments disagree by $\sim$8 seconds (out of a lifetime of $\sim$880~s), which corresponds to a nearly 4 $\sigma$ difference. The most likely explanation for this discrepancy is an underestimation of systematic effects. It is notable that this discrepancy is large enough that it is the dominant uncertainty in the prediction of the primordial helium/hydrogen abundance ratio.

It is essential to improve the reliability of both beam and bottle measurements of the neutron lifetime, as they have independent and very different systematic effects. We have outlined a path to achieve this that is affordable, is technically feasible, and engages a committed community of researchers. A new magneto-gravitational neutron bottle experiment at LANL (UCN$\tau$), which eliminates material wall interactions and mitigates other systematic effects, is capable of achieving an $\sim$1 second accuracy. At NIST, the latest version of the neutron beam experiment is taking advantage of a recent advance in the accurate determination of neutron flux and should also provide a measurement with an uncertainty of $\sim$1 second. The realization of two systematically independent experiments at the $\sim$1 second level will provide a robust determination of the neutron lifetime that is sufficiently accurate for cosmology. This result should be available in the next 3-4 years.

A measurement with an accuracy of few tenths of a second is required to substantively probe physics beyond the standard model at an energy scale beyond the reach of the LHC.  This will test the unitarity of the CKM matrix at a level that approaches 10$^{-4}$. Important recent technological advances by the US neutron lifetime community have established a clear technical path toward neutron lifetime measurements with this uncertainty. This is an important opportunity, and its realization will project the US into a position of clear leadership in this field. This further work will require modest capital and operational resources.  Both of these projects are being carried out by mature collaborations that include university and national laboratory involvement, and both projects include several senior researchers who intend to devote a majority of their research effort to these projects.  

While these projects are distinct in their experimental methodology, they both exist within the context of the national neutron nuclear physics program.  The active communication between these efforts strengthens both programs as well as the larger fundamental neutron physics community. In summary, the current program of US beam and magnetic bottle experiments should lead to a lifetime that can resolve the current discrepancy and provide needed input for cosmology.  This program also lays out a path to a result with sufficient accuracy to be an important probe of physics beyond the standard model.

\section{Introduction}
This white paper was prepared at a workshop held at the Amherst Center for Fundamental Interactions at the University of Massachusetts Amherst in September of 2014 and follows from a workshop held in Santa Fe in 2012~\cite{Seestrom2012}.  These meetings were held in response to the substantial progress that has been made by the US fundamental neutron physics community in the technology of both beam and bottle experiments.
\section{The neutron lifetime in cosmology}

The standard cosmology predicts the original composition of the
Universe with high precision.  The helium abundance in particular is
predicted with a precision of better than 1\%, with the neutron
lifetime as the limiting uncertainty \cite{planck,iocco09}.  This
parameter is used to fix coupling constants for all weak processes
interconverting neutrons and protons in the big bang, and these in
turn determine the supply of neutrons available to make helium.  Other
cross sections are negligible sources of uncertainty in the helium
prediction.

The precise predictions of BBN allow stringent tests of the standard
cosmology, using astronomical abundance measurements that have been
improving in recent years.  Although the observational error on the
helium abundance remains significantly larger than the error on its
prediction, the current discrepancy between neutron lifetime
measurements is large enough to skew interpretation of the data
noticeably.  A
neutron lifetime accurate to 1 second would reduce the uncertainty of
the prediction to about the minimum achievable using standard model
weak physics and ensure that it remains an order of magnitude below
foreseeable improvements in observational uncertainties.

There is now resurgent interest in the contribution of relativistic
particles beyond the standard model (like sterile neutrinos) to the
expansion rate of the Universe at early times.  For decades, the BBN
helium abundance was the only probe of this effect.  The standard
model predicts an ``effective number of neutrino species''
$N_\mathrm{eff}= 3.046$ \cite{mangano02}, expressing the expansion
rate in terms of the number of neutrino species given an idealized
thermal history.  (Deviations from the expected expansion rate need
not arise from thermally-populated light particle species, so
$N_\mathrm{eff}$ is more a useful parameterization than a literal
number of light particles.)  Recent observations of the cosmic
microwave background indicate that $N_\mathrm{eff}= 3.30\pm 0.27$
\cite{planck}, though there is some evidence for larger values.
Additional physical effects will constrain $N_\mathrm{eff}$ further as
the microwave background data improve.  Applying BBN to current
abundance data, one obtains $N_\mathrm{eff} = 3.56\pm 0.23$ from the
currently recommended neutron lifetime (updating Ref.~\cite{nollett14}
to a new deuterium abundance).  BBN provides an important consistency
check, particularly since it probes much earlier times in the history
of the Universe.  The present discrepancy between methods of measuring
the neutron lifetime corresponds to a shift of $N_\mathrm{eff}$ by
about 0.12 \cite{nollett11}; this is less than the size of the
observational errors, but not negligible.

\section{Neutron Lifetime and CKM Unitarity: Probing BSM Physics}

Tests of the unitarity of the Cabibbo-Kobayashi-Maskawa matrix provide one of the most powerful low-energy probes of possible physics beyond the standard model (BSM). Of particular interest to nuclear physics are tests of the first row unitarity constraint: $|V_\mathrm{ud}|^2+|V_\mathrm{us}|^2+|V_\mathrm{ub}|^2=1$. Experimentally, one finds 
\begin{equation}
\label{eq:ckm1}
\Delta_\mathrm{CKM} \equiv \left(|V_\mathrm{ud}|^2+|V_\mathrm{us}|^2+|V_\mathrm{ub}|^2\right)_\mathrm{exp}-1=-0.0001\pm 0.0006,
\end{equation}
where the uncertainty is shared roughly equally between the uncertainties in $V_\mathrm{ud}$ and $V_\mathrm{us}$~\cite{Antonelli10}. At the 95\% confidence level, this situation represents agreement with the standard model at the part per mil level, roughy commensurate with the precision obtained in the neutral current sector with precise measurements of the $Z$-boson properties at LEP and SLC. The corresponding severity of constraints implied by Eq.~(\ref{eq:ckm1}) and the $Z$-pole studies are, thus, comparable.
\
The theoretical motivation for reducing the uncertainty in $\Delta_\mathrm{CKM}$ to the $10^{-4}$ level is strong. BSM interactions such as weak scale supersymmetry~\cite{Bauman12} may be realized in nature but difficult to discover at the Large Hadron Collider if the spectrum is compressed. In this case, the loops involving supersymmetric particles could  lead to apparent shifts in $\Delta_\mathrm{CKM}$ at the $10^{-3}-10^{-4}$ level for superpartner masses in the several hundred GeV range, even if such particles are not observed at the LHC. Similarly, tree-level exchanges of new, as yet unseen, particles such as leptoquarks could generate similar shifts in $\Delta_\mathrm{CKM}$ for masses at the TeV scale or higher.  

Achieving $10^{-4}$ sensitivity with $\Delta_\mathrm{CKM}$ will require experimental and theoretical advances on several fronts. Improvements in the determination of $V_\mathrm{us}$ will be needed. The error in the extraction of $V_\mathrm{us}$ from $K_{\ell 3}$ decays is dominated by theoretical uncertainties in the kaon form factor, $f_K^+(0)$. A similar situation holds for the determination using $K_{\ell 2}$ decays, where the largest error arises from the value of decay constants $f_K/f_\pi$. Achieving a $10^{-4}$ level determination of $\Delta_\mathrm{CKM}$ will, thus, require improved theoretical computations of these quantities. Doing so represents an important challenge for lattice QCD efforts.

At present, the most precise determination of $V_\mathrm{ud}$ is obtained from superallowed nuclear $\beta$-decays~\cite{hardy13}, combined with theoretical computations of the standard model electroweak radiative corrections. Further reductions in the SM radiative correction uncertainties associated with the $W\gamma$ box graphs will be essential. Theoretical nuclear structure uncertainties suggest that a $10^{-4}$ determination of $|V_\mathrm{ud}|$ from superallowed decays will be challenging even with the advent of improved SM radiative correction calculations. Consequently, a  determination of $|V_\mathrm{ud}|$ from neutron decay~\cite{Cirigliano:13} provides a promising alternative path to $10^{-4}$ precision. Achieving this goal will require a robust measurement of $\tau_n$ with 0.1 s uncertainty and an order-of-magnitude reduction in the uncertainty of $g_V/g_A$ as obtained from neutron decay correlation studies.

\section{Experimental Status of the Neutron Lifetime}

The neutron lifetime is measured using either beams of cold neutrons or storage of ultracold neutrons in material or magnetic bottles.  Both techniques have similar statistical power, yet they offer completely different sets of systematic effects.  Given the difficulty in accurately measuring the neutron lifetime, comparison between bottle and beam experiments is essential to having a reliable value of the neutron lifetime.

In beam experiments, it is necessary to account for the number of neutrons in a well-characterized length of beam and for the number of decays occurring.  The number of neutrons is best counted with a detector whose response to neutrons is inversely proportional to the velocity of the neutron. Neutron decays can be monitored by counting either the decay protons, the decay electrons, or both in coincidence.  The neutron lifetime is proportional to the ratio of the average number of neutrons to their decay rate.

The most precise beam experiment used a quasi-Penning trap to capture decay protons.  The 2005 result from this NIST-based experiment was 886.3$\pm$3.4~\cite{Nic05} s with the largest uncertainty due to neutron counting.  Recently, the neutron detector was re-calibrated to 0.06\% relative uncertainty, making it possible to revise this result to 887.7$\pm$2.3~s~\cite{Yue13}.  On the international level, there is a new beam method effort currently under development at JPARC in Japan that uses a pulsed neutron beam and a $^3$He TPC.
In bottle experiments, ultracold neutrons (UCN) have been confined in material or magnetic bottles, using either the Fermi potential of the walls or regions of high magnetic field to confine the neutrons.  Gravity can be used to close the bottle at the top.  To enable a precise measurement of the neutron decay rate, great care is taken in experiments using material bottles to keep additional losses at a minimum and to characterize any residual losses arising from imperfect wall reflection. Pioneering experiments used strategies such as extrapolation to zero wall loss by combining data for different trap geometries and scaling storage time intervals to ensure reliability of the extrapolation. In some experiments, additional information on wall losses is obtained from the measurement of neutrons up-scattered by the walls. In magneto-gravitational bottles, losses due to materials interactions are completely eliminated, along with their corresponding systematic uncertainties. These bottles are subject to losses due to spin flip of the trapped neutrons; but these losses have been shown to be small if B-field zeros are avoided~\cite{steyerl12,Seestrom2012}.
\begin{figure}
\begin{center}
\includegraphics[width=0.5\textwidth, angle=270]{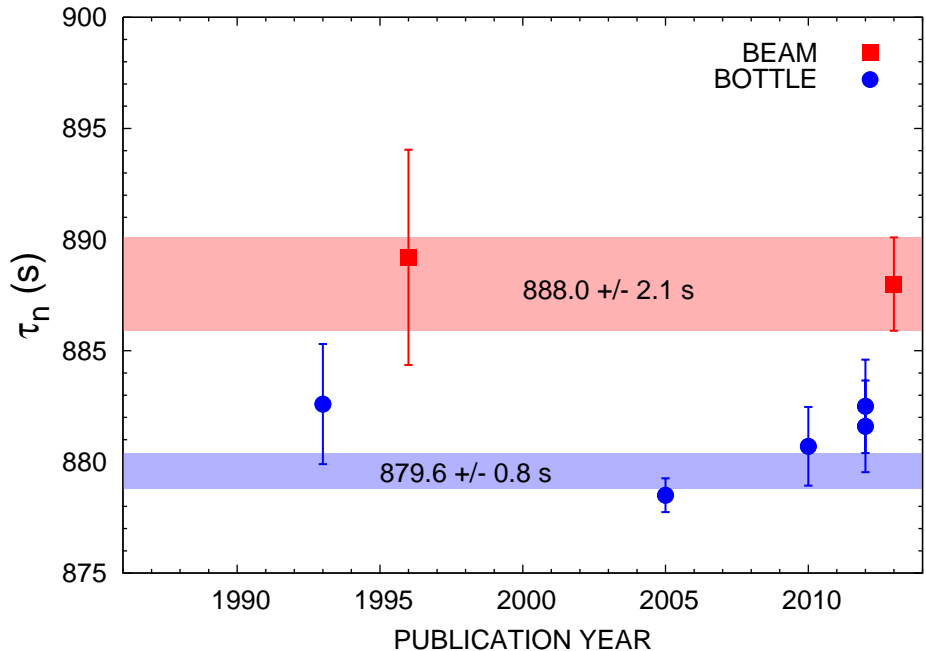}
\caption{Published results from beam and bottle measurements of the free neutron lifetime currently used by the PDG.  Separate averages are shown.}
\label{fig:status}
\end{center}
\end{figure}

The current Particle Data Group (PDG) mean value for the neutron lifetime is based on two beam experiments by the group at NIST and five material bottle measurements using liquid or frozen fluorinated oil as a low-loss wall material. The two beam experiments in the PDG selection, averaged separately, yield the value 888.0$\pm$2.1~s for the lifetime, higher than the current average of 879.6$\pm$0.8~s for bottle experiments.  The difference between beam and bottle experiments requires urgent attention (Fig.~\ref{fig:status}).

In addition to the U.S. experiment UCN$\tau$ (discussed in Section~\ref{bottle_sec}) there are four either planned or ongoing magnetic trap experiments in Europe.  The most mature of these is the Ezhov~\cite{ezhov_ref} experiment, with a larger volume trap under construction and available at the end of 2014 at the earliest.  The other experiments, all in the construction phase, are HOPE~\cite{leung09,Seestrom2012} at the ILL, PENeLOPE~\cite{matern09,Seestrom2012} based at Technical University Munich, and the Mainz experiment~\cite{mainz_ref}.  Of these efforts, HOPE and PENeLOPE plan to implement more than one method of monitoring UCN populations in the trap, and only PENeLOPE is comparable in volume to UCN$\tau$.  However, UCN$\tau$'s asymmetric construction and internal decay monitoring are unique.


\section{The Beam Neutron Lifetime Program}
\label{S:BL3}
The Sussex-ILL-NIST beam neutron lifetime method counts neutron decay protons trapped in a quasi-Penning trap. This is a mature program that began more than 30 years ago and many systematic effects have been thoroughly studied and are very well understood.  Much of the research and development has already been done and potential improvements to the method and apparatus have been identified. Until recently, the most precise result from this program was $\tau_n = 886.3\pm3.4$ s \cite{Nic05}, carried out at the NIST Center for Neutron Research with publication in 2005. The neutron counter efficiency was the largest source of uncertainty, in particular from the areal density of the $^6$LiF neutron absorbing foil and the $^6$Li$(n,\alpha)$ cross section. A significant reduction in this uncertainty required an independent absolute neutron flux calibration of the detector, which was successfully completed in 2013 using a $^{10}$B alpha-gamma spectrometer \cite{Yue13}. It enabled an update to the 2005 result and a reduction in overall uncertainty to $\tau_n = 887.7\pm2.3$ s. This important development has opened the door for reducing other uncertainties in order to reach the 1 s precision level and below.
\begin{figure}[hb]
\centering
\includegraphics[width = 6.5in]{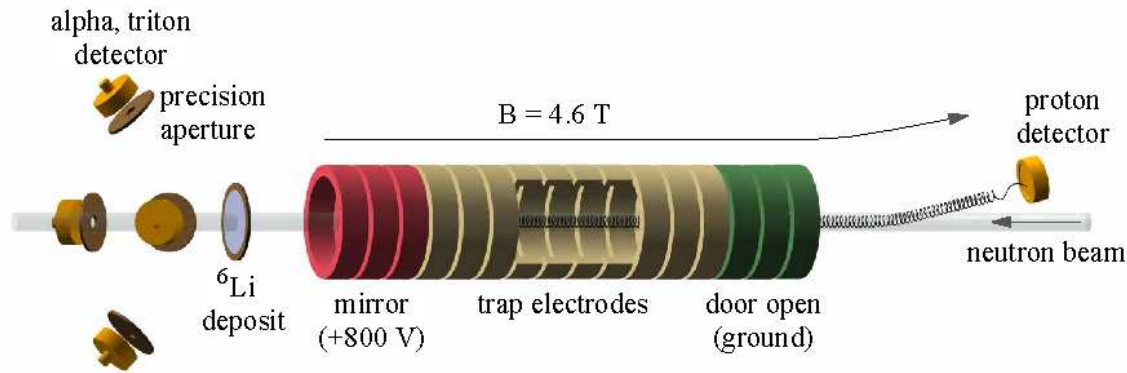}
\vspace{-0.2in}
\caption{\label{F:tNscheme} Schematic of the beam neutron lifetime experiment using the Sussex-ILL-NIST method. The neutron beam passes through
a quasi-Penning trap. Decay protons are trapped by the elevated door and mirror electrode potentials, and counted periodically by lowering the door to ground. Neutrons are counted by detecting the alphas and tritons from the $^6$Li$(n,\alpha)$ reaction in a thin $^6$LiF deposit.}
\end{figure}
\par
The next step (called ``BL2'') in this program is to repeat the experiment, using the existing apparatus with key improvements in neutron flux and trap linearity. This effort, which will run at NIST in 2015--2016, aims for a total uncertainty of $\sim$1~s in the neutron lifetime. In parallel with this effort, it is proposed to design and construct an entirely new apparatus (called ``BL3'') capable of reaching a significantly smaller experimental uncertainty. Figure \ref{F:tNscheme} shows the general scheme of the method that was used for the 2005 experiment and applies to BL2 and BL3 as well. The BL3 experiment has two goals:
\begin{enumerate}
\item Explore, improve, and test all known systematic effects to the $10^{-4}$ level. The current 8 s disagreement between the beam and bottle methods
is the biggest impediment to improving the precision of the accepted value of the neutron lifetime. This must be addressed by verifying that no important systematics in the beam method have been overlooked or wrongly estimated.
\item Reduce the final uncertainty of the beam method neutron lifetime to well below 1 second.
\end{enumerate}
BL3 requires an entirely new and significantly larger apparatus to obtain the desired increase in proton counting statistics. This also presents an opportunity to enact a number of systematic improvements to the method. Key design features of the new experiment are:
\begin{itemize}
\item The neutron beam diameter will increase from 7 mm to 35 mm and the new trap will be larger in diameter and twice as long. This, along with the expected increase in neutron flux at the new NIST NG-C beamline, will result in a proton counting rate 200 times higher than in the 2005 experiment.
\item The new superconducting magnet will be larger and the field much more uniform in the proton trapping region: $\Delta B / B < 10^{-3}$ for an expected trap nonlinearity correction of $<$ 0.1 s. A set of trim coils will be added to adjust the uniformity and to perform systematic tests by varying the magnetic field profile.
\item A segmented 10-cm diameter silicon detector, with a very thin dead layer compared to the previous proton detector, will be used. It will be similar to the detectors recently developed at Los Alamos National Lab for the Nab and UCNB experiments.
\item The proton detector will be located in a region of magnetic field gradient and have the ability to be translated along the axis to realize a variable magnetic field expansion. This will enable a large reduction in the proton backscattering correction and uncertainty.
\item A new  $^{10}$B alpha-gamma spectrometer will be designed and built to accommodate the larger neutron beam and allow systematic improvements to achieve an improved precision of $1\times10^{-4}$.
\item A second, independent absolute neutron counting method will be used to verify the neutron flux calibration. Two different schemes: a $^3$He gas scintillator, and a low-temperature neutron calorimeter, are currently under development for this.
\item A neutron time of flight spectrometer will be integrated into the design. This will enable a precision measurement of the neutron velocity spectrum needed to improve the uncertainty on corrections due to neutron absorption and scattering in the neutron flux monitor ($^{6}$Li and Si).
\end{itemize}
BL3 will use a scaled-up version of the existing proven and well understood apparatus, with a number of incremental technical improvements, so no high-risk research and development is needed. The collaborating institutions include Tulane University (F.~Wietfeldt), University of Tennessee (N.~Fomin, G.~Greene), NIST (M.~Dewey, A.~Yue), and Indiana University. The identified senior personnel expect to devote the majority of their research effort to this project. The experiment will run at the NG-C fundamental physics end position at the NIST Center for Neutron Research.

\section{The Magneto-Gravitational Neutron Bottle Program}  
\label{bottle_sec}
A research and development program is underway to study techniques for a next-generation experiment aimed at significantly better than 1 second precision and to design an experiment capable of such precision.
A collaboration, named UCN$\tau$, has been formed and plans to use an existing prototype magneto-gravitational trap for UCNs to study systematic effects and, as an intermediate goal during the course of this R\&D program, measure the neutron lifetime to $\sim$1 second precision, comparable to the current world-average precision.

A number of features have been identified as desirable to perform a high-precision neutron lifetime measurement using the bottle technique~\cite{Seestrom2012}.

\begin{itemize}
\item 
An intrinsically low loss rate and long trap lifetime is necessary to improve the accuracy with which systematic corrections must be made. 
This can be achieved with the
elimination of neutron-material interactions by using magnetic fields and gravity, instead of the material walls of the recent bottle experiments, to confine the neutrons. 

\item Rapid evolution and mixing of the phase space of the stored neutron population has multiple benefits: it allows the fast (compared to the neutron lifetime) removal of quasi-bound (long-lived, untrapped) neutrons from the bottle, and it permits the detection of surviving neutrons with efficiency independent of storage time. 
The desired ergodic UCN dynamics can be achieved through careful design of the trap geometry and detailed surface profiles.

\item A detection scheme that is insensitive to the phase space evolution of the trapped neutrons will provide further insurance against bias introduced by any residual temporal change in the dynamics of the trapped UCNs.

\item The statistical reach to make a measurement at the ultimate goal precision every few days will enable the data-driven study of systematic effects.
This statistical power can be achieved in bottle experiments by a combination of large volume, high UCN source density, and high neutron detection efficiency.

\end{itemize}

\noindent The prototype apparatus, based on the magneto-gravitational UCN trap concept detailed in Ref~\cite{Walstrom200982}, realizes these desired features.
The asymmetric magnetic field profile of the bottle is specifically designed to enable fast removal (``cleaning'')
of quasi-bound orbits via rapid mixing of phase space and to be free of magnetic-field zeros, thereby minimizing (or eliminating) systematic corrections that would otherwise be required; 
ideally, the storage lifetime in this trap after a short cleaning period is precisely the free neutron beta decay lifetime.
The bottle volume is large, about 600~liters, such that the UCN density now achieved at the LANL UCN source is sufficient 
to trap $\sim$10$^5$ neutrons per fill, enough statistics for a 1 second measurement on a time-scale of
$\sim$1~day.

The experiment utilizes an {\it in situ} neutron detector based on neutron absorption on a vanadium foil.  
The vanadium foil ($^{51}$V) is lowered into the UCN trap after a defined holding period to absorb the surviving UCNs, and then activity is measured by detecting the $\beta$ and $\gamma$ emitted in $^{52}$V decay.  The $\beta$-$\gamma$ coincidence is used to suppress the background.
The V-foil method is much less sensitive than previous measurement techniques to a possible bias through the coupling of non-uniform detector efficiency to phase-space evolution of the stored UCNs or, if the remaining UCNs are counted by draining the bottle into an external detector, draining time variations.

A prototype apparatus has been constructed using LANL LDRD and IU NSF funds.
It was used to complete a proof-of-principle magnetic storage measurement~\cite{Salvat:2013gpa} at the Los Alamos UCN facility.
It has further been used to demonstrate a very long trap lifetime ($>$ 10~h), an {\it in-situ} detector efficiency of 50\%, and an initial neutron population in the bottle of $\approx$10$^5$ per fill, together allowing a 1 second statistical measurement of the neutron storage lifetime in 48 hours.    
Additional desirable features of this apparatus are the room temperature storage bottle (which allows rapid experimental turn-around) and the open-top geometry (which allows easy access for prototype detectors, neutron-spectrum cleaners, and other devices).
\begin{figure}[h!]
\centering
(a)\includegraphics[width=0.4\textwidth]{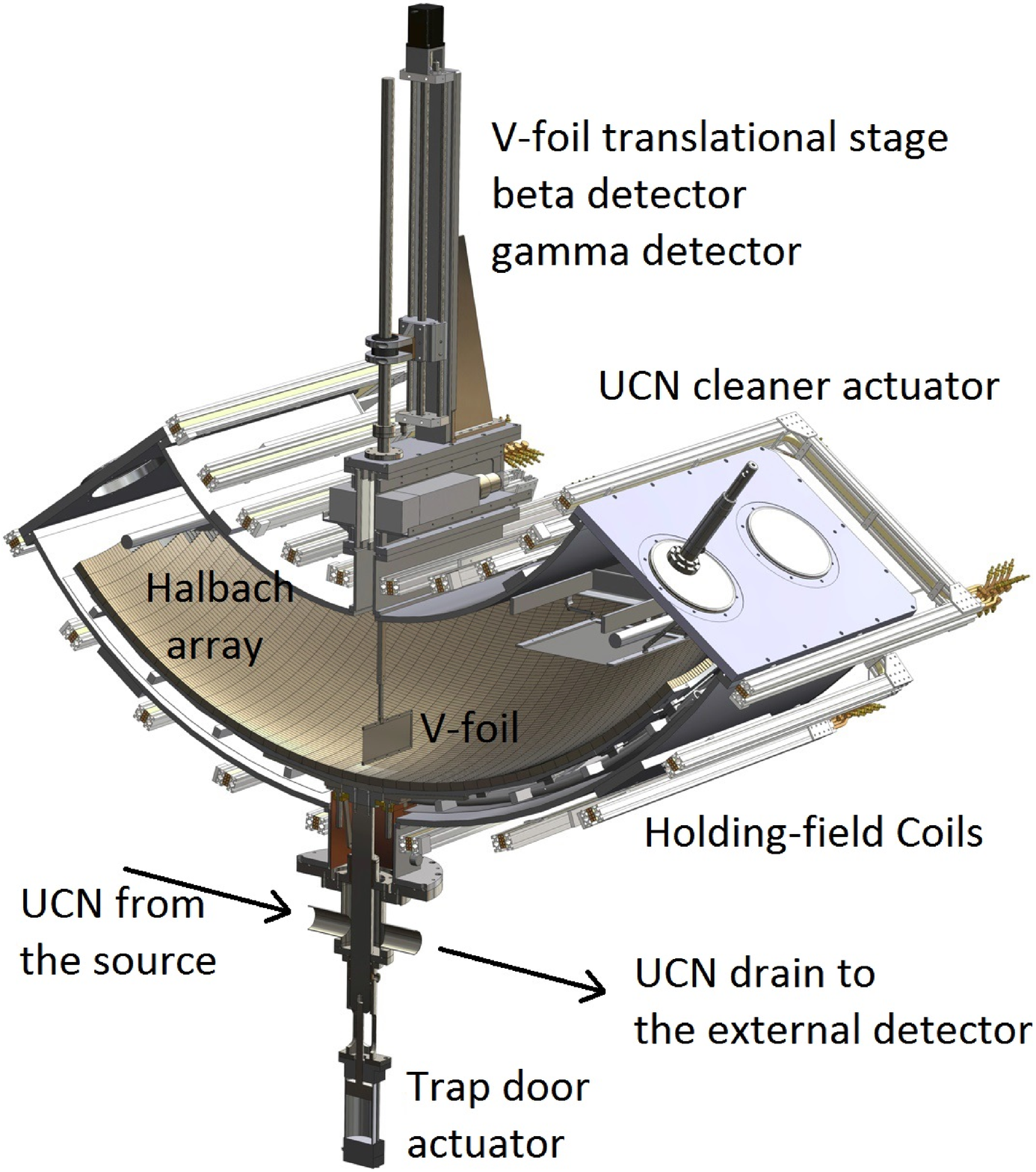}
(b)\includegraphics[width=0.3\textwidth]{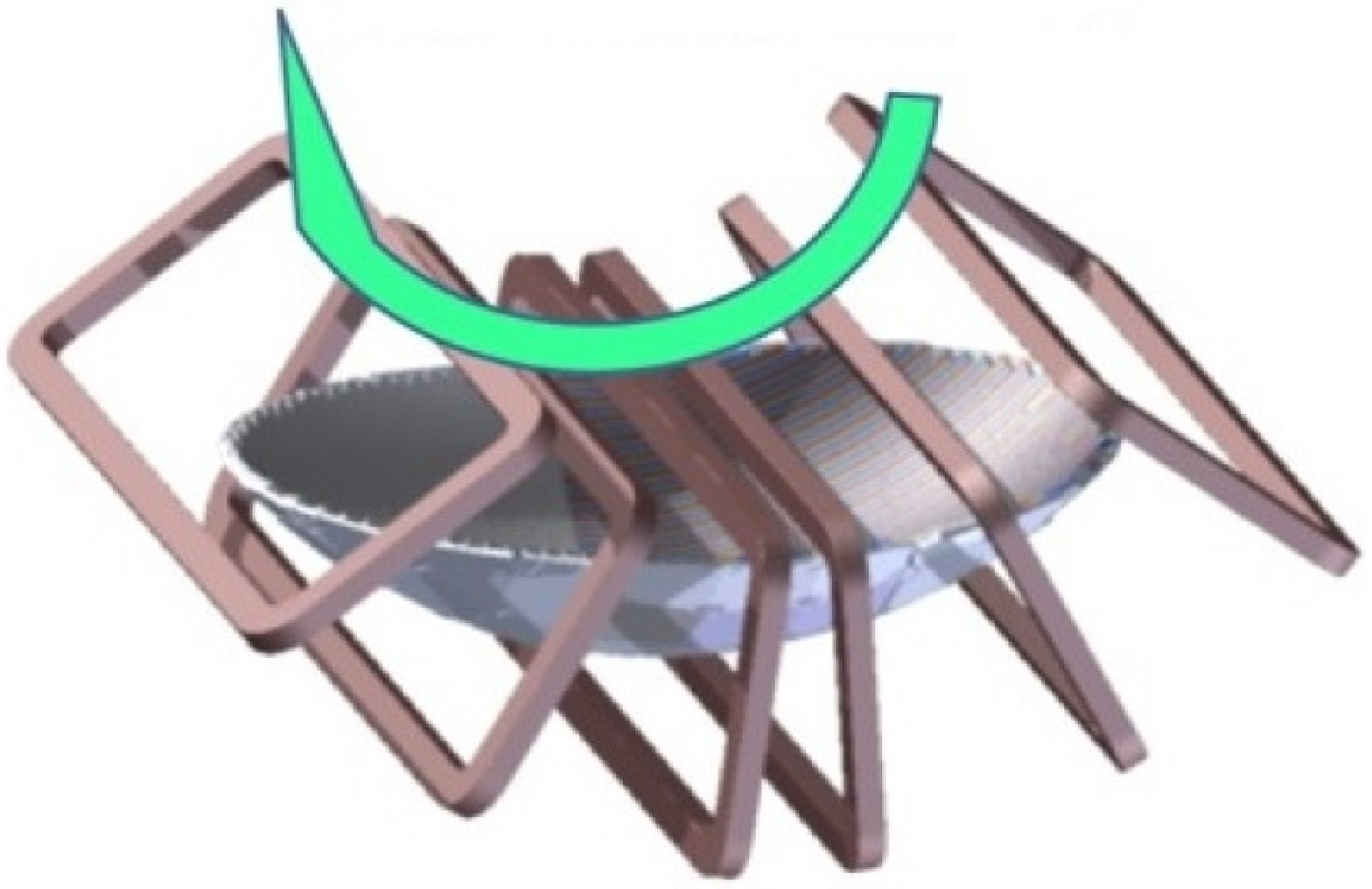}
\caption{(a) A cut-away view of the experimental apparatus. (b) The permanent magnet array (provide the trapping field) and the electromagnetic coils (provide the spin holding field). 
The two fields are perpendicular everywhere inside the trap volume, eliminating any field zero that could lead to neutron spin flip. 
}
\label{fig:trap2}
\end{figure}
The UCN$\tau$ collaboration consists of 40 scientists from 10 institutions.
The senior scientists, from Indiana University (Liu), Los Alamos National Laboratory (Saunders, Clayton, Makela, 
Morris, Seestrom), Tennessee Technical University (Holley), and DePauw University (Komives), 
will devote the majority of their research time to this effort when the program is funded.

Over the next 3 years, the collaboration will use this prototype 1) for an ongoing series of systematic studies at the 1 second precision level and 2) to investigate the requirements to achieve a total precision well below 1 second. 
A 2012 workshop held at Santa Fe resulted in the endorsement of the approach adopted by UCN$\tau$ as the most promising path to a sub second bottle lifetime measurement.

To achieve a total precision of well below 1 second, a series of technical hurdles will need to be overcome. 
A thorough study of chaos theory and its applications to neutron orbits in the storage bottle, possibly resulting in a new, optimized bottle design, has begun~\cite{Seestrom2012}.
The development of high-efficiency neutron counting is progressing rapidly.    
An upgrade to the LANL UCN facility, which will provide the necessary UCN density to achieve rapid sub second statistical reach, is funded and underway.  The competing European measurements take advantage of the ability to eliminate material interactions that is inherent in magnetic bottle experiments. Of these experiments, HOPE and PENeLOPE plan to implement more than one method of monitoring UCN populations in the trap, and only PENeLOPE is comparable in volume to UCN$\tau$.  However, UCN$\tau$'s asymmetric construction and internal decay monitoring are unique.

The low non-$\beta$-decay loss rate and high statistical power 
will enable data-driven studies of subtle systematic effects.
Here, several lifetime measurements are performed under varying conditions,for example with intentionally increased numbers of quasi-bound UCNs loaded into the trap.  Another example is the effect of possible phase-space evolution during the UCN storage period, which can be studied by comparing measured lifetimes with varied storage times.
The result of the present (2--3 year) R\&D program will be a fully-evaluated prototype UCN$\tau$ apparatus, much-improved understanding of relevant systematic effects, and ultimately a design for a next-generation magneto-gravitational trap experiment with goal precision well below than 1~second.

\section{Summary}
A strong US experimental program to measure the neutron lifetime, including at least two experiments with independent systematic effects, is needed.  The first set of measurements will be done at the 1~second uncertainty level  to resolve the present discrepancy between existing experimental methods.  Follow-up experiments would be done at an uncertainty level well below 1~second to probe for physics beyond the Standard Model.  The present discrepancy has consequences across physics, in particular providing the largest source of uncertainty in BBN predictions.  Two experimental programs in the US, one at NIST using a well-established and mature beam-type experimental method and the other at LANL using a novel asymmetric magneto-gravitational bottle-type method, are well positioned to make 1~second precision measurements in the next few years, followed by clear paths to independent measurements with total uncertainties well below 1 second.  Funding both of these efforts will result in a robust value of the neutron lifetime capable of probing new physics at and beyond the LHC level.

\bibliographystyle{h-physrev3.bst}


\end{document}